\documentstyle[multicol,aps,prl,psfig]{revtex}

\renewcommand{\narrowtext}{\begin{multicols}{2} \global\columnwidth20.5pc}
\renewcommand{\widetext}{\end{multicols} \global\columnwidth42.5pc}
\def\ga{\gamma}
\def\de{\delta}

\def\ve{\varepsilon}

\def\et{\eta}

\def\la{\lambda}

\def\si{\sigma}

\def\ph{\phi}

\def\ch{\chi}
\def\ps{\psi}

\def\Ga{\Gamma}

\def\mn{{\mu\nu}}
\def\cl{{\cal L}}
\def\fr#1#2{{{#1} \over {#2}}}
\def\prt{\partial}

\def\abs#1{\left|{#1}\right|}

\def\half{{\textstyle{1\over 2}}}

\def\frac#1#2{{\textstyle{{#1}\over {#2}}}}

\def\lsim{\mathrel{\rlap{\lower4pt\hbox{\hskip1pt$\sim$}}
    \raise1pt\hbox{$<$}}}
\def\gsim{\mathrel{\rlap{\lower4pt\hbox{\hskip1pt$\sim$}}
    \raise1pt\hbox{$>$}}}
\def\sqr#1#2{{\vcenter{\vbox{\hrule height.#2pt
         \hbox{\vrule width.#2pt height#1pt \kern#1pt
         \vrule width.#2pt}
         \hrule height.#2pt}}}}

\def\ol#1{\overline{#1}}
\def\ce{{\cal E}}
\def\co{{\cal O}}
\def\cb{{\cal B}}
\def\cp{{\cal P}}
\def\cv{{\cal V}}
\def\a{$a_\mu$}
\def\b{$b_\mu$}
\def\c{$c_{\mu\nu}$}
\def\d{$d_{\mu\nu}$}
\def\e{$e_\mu$}
\def\f{$f_\mu$}
\def\g{$g_{\la\mu\nu}$}
\def\H{$H_{\mu\nu}$}

\def\da#1{{#1}^{\dag}} 
\def\lrvec#1{ \stackrel{\leftrightarrow}{#1} }
\def\ad{ {\rm ad} }

\def\ea#1{ \exp[\ad(iS_{#1})] }

\def\norm#1{\left\|{#1}\right\|}

\newcommand{\beq}{\begin{equation}}
\newcommand{\eeq}{\end{equation}}
\newcommand{\bea}{\begin{eqnarray}}
\newcommand{\eea}{\end{eqnarray}}
\newcommand{\rf}[1]{(\ref{#1})}

\newcommand{\bdm}{\begin{displaymath}}
\newcommand{\edm}{\end{displaymath}}

\begin{document}

\title{Nonrelativistic Quantum Hamiltonian for Lorentz Violation}
\author{V.\ Alan Kosteleck\'y and Charles D.\ Lane}
\address{Physics Department, Indiana University, 
          Bloomington, IN 47405, U.S.A.}
\date{IUHET 404, April 1999; accepted for publication in J.\ Math.\ Phys.}
\maketitle

\vskip 0.1 truecm

\begin{abstract}
A method is presented for deriving 
the nonrelativistic quantum hamiltonian 
of a free massive fermion from the relativistic lagrangian
of the Lorentz-violating standard-model extension.
It permits the extraction of terms at arbitrary order 
in a Foldy-Wouthuysen expansion in inverse powers of the mass.
The quantum particle hamiltonian is obtained
and its nonrelativistic limit is given explicitly to third order.

\end{abstract}

\section{Introduction}
\label{intro}

Establishing the physical relevance of a lagrangian
in relativistic quantum field theory
often requires a determination 
of its nonrelativistic content.
The Foldy-Wouthuysen (FW) transformation \cite{fw} 
provides a systematic approach to understanding 
the low-energy effects of certain theories.
Given the relativistic quantum hamiltonian 
for a theory of massive four-component fermions,
the nonrelativistic quantum hamiltonian 
for the corresponding two-component particle
can be derived in an expansion in inverse powers
of the fermion mass.

In this work,
we use generalized FW methods to 
investigate the quantum particle hamiltonian 
that describes the physics of a free massive two-component fermion 
emerging from the relativistic lagrangian
of the Lorentz-violating standard-model extension
\cite{cksm}.
This standard-model extension is based on the idea
of spontaneous Lorentz breaking in an underlying theory
\cite{kps}
and has been used 
for various investigations placing constraints
on possible violations of Lorentz symmetry
\cite{cksm,cpt98,bkr,gg,hd,rm,ckpv,kexpt,bexpt,bckp,cg,bkl,cptcc},
several of which depend crucially on 
the nonrelativistic physics of free massive fermions. 
In these investigations,
specific terms in the nonrelativistic hamiltonian
have been derived as needed,
but a full treatment has been lacking. 
Here,
we provide a systematic approach
that permits extraction of the relevant terms 
in the nonrelativistic hamiltonian
at arbitrary order in the FW approximation.
We obtain the quantum particle hamiltonian 
and provide explicitly the form 
of the nonrelativistic hamiltonian to third order.
Our results are directly relevant to 
recent analyses of muon and clock-comparison experiments
\cite{bkl,cptcc}
and are expected to have substantial impact on
further studies of the physical implications of
the standard-model extension.

The general form of the relativistic lagrangian
for a free spin-$\half$ Dirac fermion $\ps$ of mass $m$
in the standard-model extension
is \cite{cksm}
\beq
\cl = \frac{1}{2} i \ol{\ps} 
(\ga_\nu + c_\mn \ga^\mu + d_\mn \ga_5 \ga^\mu 
   + e_\nu + i f_\nu \ga_5
   + \half g_{\la \mu \nu} \si^{\la \mu})
\lrvec{\prt^\nu} \ps 
- \ol{\ps} 
(m + a_\mu \ga^\mu + b_\mu \ga_5 \ga^\mu 
   + \half H_\mn \si^\mn)
   \ps
\quad .
\label{lagr}
\eeq
This is a generalization of the usual relativistic lagrangian 
for a free massive Dirac fermion.
The Dirac matrices 
$\{ 1, \ga_5, \ga^\mu, \ga_5\ga^\mu, \si^\mn \}$ 
have conventional properties,
and the Minkowski metric $\et_{\mu\nu}$ has signature $-2$.
The parameters \a, \b, \c, \d, \e, \f, \g, and \H\
control the extent of Lorentz violation in the theory.
In a given observer inertial frame,
they can be regarded as fixed real Lorentz vectors or tensors.
Note that \H\ can be taken as antisymmetric, 
\c\ and \d\ as traceless,
and \g\ as antisymmetric in the first two indices.
Since Lorentz symmetry is known to be valid to high precision,
any nonzero parameters in nature would need to be minuscule.
We therefore restrict our attention in this work 
to terms linear in these parameters.

In the next section,
the relativistic particle-antiparticle hamiltonian $H$ 
corresponding to the lagrangian \rf{lagr} is obtained.
Some basic information about
our procedure for extracting 
its FW form is discussed in 
section \ref{sequence},
together with our definition of 
the relevant FW sequence.
Features of this sequence are derived 
in section \ref{calculation},
and the quantum particle hamiltonian
and its nonrelativistic limit to third order 
are explicitly presented in section \ref{nrham}.

\section{Relativistic Quantum Hamiltonian}
\label{relham}

The first step in deriving low-energy effects
of the Lorentz-violating terms
is to obtain the relativistic hamiltonian $H$ associated 
with the lagrangian \rf{lagr}.
However, 
methods for direct construction of $H$ 
are inadequate 
because Eq.\ \rf{lagr} contains couplings 
involving time derivatives.
For example,
applying the Euler-Lagrange equations to $\cl$
and solving for $H$ 
from the equation $i\prt_0 \ps=H\ps$
results in a non-hermitian hamiltonian
and a corresponding nonunitary time evolution.

One method 
of bypassing this technical difficulty
is to perform a field redefinition $\ps = A \ch$ 
in the lagrangian
\cite{bkr},
with $A$ chosen such that the dependence of the lagrangian
on $\prt_0\ch$ is just that of the usual Dirac lagrangian.
Then,
the wave function associated with $\ch$
evolves conventionally in time.
The field redefinition leaves unchanged the physics,
while it causes the time-derivative couplings
to be replaced by extra terms in the lagrangian.

To implement this idea,
we write the lagrangian \rf{lagr} in the forms
\bea
\cl &\equiv & \frac{1}{2} i \ol{\ps} \Ga_\nu \lrvec{\prt^\nu} \ps 
- \ol{\ps} M \ps
\nonumber\\ 
&=& \frac{1}{2}i\ol{\ch} \ga_0 \lrvec{\prt^0} \ch
   +\frac{1}{2}i\ol{\ch}(\ol{A}\Ga_j A) \lrvec{\prt^j} \ch 
   - \ol{\ch}(\ol{A}MA)\ch
\quad ,
\label{fr1}
\eea
where $\Ga_\nu$ and $M$ are defined 
according to the correspondence with Eq.\ \rf{lagr},
and $\ol{\ps}=\ol{\ch}\ol{A}$ with
$\ol{A}:=\ga^0\da{A}\ga^0$.
In the second expression
the Lorentz indices are separated into timelike and spacelike
cartesian components, $\mu \equiv 0$ and $j = 1,2,3$,
with summation on repeated indices understood.

The choice 
\beq
A = 1- \half \ga^0 (\Ga_0 - \ga_0) 
\quad , \qquad
\ol{A} = 1 - \half (\Ga_0 -\ga_0) \ga^0
\quad
\label{AAbar}
\eeq
implements the equality \rf{fr1}
to linear order in the parameters for Lorentz violation.
Derivation of the relativistic hamiltonian $H$ 
can then proceed through the Euler-Lagrange equations,
which take the form of a modified Dirac equation:
\beq
( i \ol{A} \Ga_\mu A \prt^\mu - \ol{A} M A)\ch = 0
\quad .
\label{Deqn}
\eeq
We find 
\beq
H = - \ga^0 \ol{A} \Ga_j A p^j + \ga^0 \ol{A} M A
\quad ,
\label{relham1}
\eeq
where the three-momentum of the particle is denoted $p_j$,
and $H$ obeys the equation $i\prt_0\ch=H\ch$. 

Explicitly,
the relativistic hamiltonian can be written 
\beq
H=m (\ga^0 + \cp_0 + \co_0 + \ce_0 )
\quad ,
\label{relham2}
\eeq
where 
\bea
m \cp_0 &:=& -p_j \ga^0 \ga^j \quad , 
  \nonumber \\
m \co_0 &:=& 
  [ -b_0 + (d_{0j}+d_{j0})p^j ] \ga_5 
     + [ a_j - (c_{jk}-c_{00}\et_{jk})p^k ] \ga^0 \ga^j  
  +i f_j p^j \ga_5 \ga^0 
     + i [ H_{0j} + (g_{j0k}+g_{jk0})p^k ] 
     \ga^j \quad  ,
  \nonumber \\
m \ce_0 &:=&
  [ a_0 - (c_{0j}+c_{j0})p^j - me_0 ] 
    +[ -b_j + (d_{jk}-d_{00}\et_{jk})p^k 
       - \half m \ve^{klm}\et_{jm} g_{kl0} ] \ga_5 \ga^0 \ga^j  
  \nonumber \\
 && -[ mc_{00} + e_j p^j ] \ga^0  
    -[ \half \ve^{klm}\et_{jm}H_{kl} + m d_{j0} 
       - \ve^{lmn}\et_{jn}(\half g_{lmk}-\et_{km}g_{l00})
     p^k] 
     \ga_5 \ga^j  
\quad .
\label{relham3}
\eea
In these expressions, 
the totally antisymmetric rotation tensor
$\ve^{jkl}$ satisfies $\ve_{123} = +1$ and  
$\ve^{jkl} = - \ve_{jkl}$.
The particular decomposition of $H$ 
into the four terms in \rf{relham2} 
is chosen for later convenience.

As an aside,
we remark that the relativistic hamiltonian 
is also readily found
if the theory \rf{lagr} is extended 
to include a minimal coupling to a U(1) gauge field $A_\mu$.
It suffices to replace the partial derivative $i\prt_\mu$
in Eq.\ \rf{lagr} with the covariant derivative
$i D_\mu := i\prt_\mu - q A_\mu$,
where $q$ is the particle charge. 
The relativistic hamiltonian then
has the same form as 
in Eqs.\ \rf{relham2} and \rf{relham3} above,
except that all occurrences of
$p_j$ must be replaced with $\pi_j:=p_j - q A_j$
and the term $qA_0$ must be added to Eq.\ \rf{relham2}.
The resulting hamiltonian is relevant,
for example,
for studies of Lorentz-violating effects in quantum electrodynamics.

\section{Definition of the FW Sequence}
\label{sequence}

In the strict nonrelativistic limit,
the lower two components of the relativistic 
wave function $\ch$ are negligible,
so the upper two components of $\ch$ suffice
to determine the nonrelativistic particle behavior. 
However, 
more generally
the Dirac equation couples the upper and lower components
of $\ch$.
The object of the Foldy-Wouthuysen procedure 
is to find a (momentum-dependent) unitary transformation 
\beq
H\mapsto\tilde{H}:=e^{iS}H e^{-iS}=\ea{}H
\quad , 
\label{nrham1}
\eeq
where ad$(X)Y:=[X,Y]$,
such that $\tilde{H}$ is $2\times 2$ block diagonal.
This therefore decouples the upper and lower components
of the FW-transformed wave function $\ph :=e^{iS} \ch$.
Requiring hermiticity of $S$ ensures
that $e^{iS}$ is unitary.
It follows that $\tilde{H}$ is hermitian
and that both hamiltonians $H$ and $\tilde H$
describe the same physics.
The FW transformation amounts to a unitary rotation 
in the Hilbert space of the free particle states
that preserves the dominance of the upper two components
of the wave function.
The quantum particle hamiltonian $h_{\rm rel}$ 
and the nonrelativistic limit $h$ we seek 
are given by the leading $2\times 2$ block of $\tilde H$. 

Solving directly for $\tilde H$ would be of interest
but is challenging in the general case.
Instead,
we present a method that allows 
approximation of $\tilde{H}$ to arbitrary accuracy
in an expansion in powers of ${|\vec p|}/m$.
The basic idea is to
apply a succession of transformations of the type \rf{nrham1}, 
chosen so that each iteration of the transformed hamiltonian
has a smaller block off-diagonal part than the previous one.
The exact FW transformation is the limit of this sequence.
Although more direct approaches can yield a low-order approximation
to $h$ without the use of our method,
the results derived here permit straightforward calculation 
of $h_{\rm rel}$ and of $h$ to any desired order. 

For definiteness in what follows,
we work within the Dirac-Pauli representation of the Dirac matrices,
for which 
$\ga^0=\left( \begin{array}{cc} 1&0\\0&-1 \end{array} \right)$
and 
$\ga^j=\left( \begin{array}{cc} 0&\si^j\\-\si^j&0 \end{array} \right)$,
where $\si^j$ are the usual Pauli matrices.
We define a matrix to be \it even \rm if it is block diagonal
and \it odd \rm if it is block off-diagonal. 
Any $4\times 4$ matrix $X$ can be uniquely
written as the sum of an even part and an odd part,
$X = {\rm Even} (X) + {\rm Odd} (X)$,
where
${\rm Odd} (X)=\half\ga^0[\ga^0,X]$ and 
${\rm Even} (X)=\half\ga^0\{\ga^0,X\}$.

We seek a sequence of FW transformations 
such that the odd part of the hamiltonian
progressively decreases in some suitable matrix norm,
such as $\norm{A}:=\max_{a,b}\{ \abs{A_{ab}} \}$ for $a, b = 1,2,3,4$.
In the remainder of this section,
an appropriate sequence $\{H_n\}$ of hamiltonians is introduced.
For each $n$, 
we also introduce a parameter $t_n$
that turns out to provide a measure of the size of ${\rm Odd}(H_n)$.
We show in section \ref{calculation}
that with our definition for the FW sequence
roughly $N$ iterations
are needed to arrive at a nonrelativistic hamiltonian
that is even to order $({|\vec p|}/m)^{(3^N-1)}$.

To start the FW sequence,
choose 
\beq
H_0=m_0 (\ga^0 + \cp_0 + \co_0 + \ce_0 )
\quad ,
\label{h0}
\eeq
where $m_0 := m$ and the terms $\cp_0$, $\co_0$, and $\ce_0$ 
are defined in Eq.\ \rf{relham3}. 
This decomposition of $H_0$ into four parts
has the following useful properties:
(i) $\cp_0$ and $\co_0$ are odd;
(ii) $\ce_0$ is even;
(iii) $\co_0$ and $\ce_0$ are first order in 
parameters for Lorentz violation,
so products of these quantities can be neglected;
and (iv) $\cp_0^2$ is proportional to the $4\times 4$ Dirac identity matrix,
with proportionality coefficient
$t_0^2={\abs{\vec{p}}^2}/{m^2}$.
We choose for the initial FW transformation
the hermitian matrix $S_0$ defined by
\beq
iS_0:=\fr{1}{2m_0}\ga^0 [{\rm Odd} (H_0)]
    = \frac{1}{2}\ga^0 (\cp_0+\co_0)
\quad .
\label{s0}
\eeq
This choice ensures that the odd part 
of $\ea{0}H_0$ is smaller than the odd part of $H_0$.

Our FW sequence is then defined iteratively by
\bea
H_{n+1}&:=& e^{iS_n} H_n e^{-iS_n}
 = \sum_{k=0}^\infty \fr{1}{k!} 
   \underbrace{ [iS_n,[iS_n, \cdots [iS_n,}_{k\ 
   \hbox{\scriptsize commutations with}\ iS_n} H_0]
   \cdots ] ] 
 = \ea{n} H_n
\quad 
\label{hjplus1}
\eea
and
\beq
iS_{n+1}:= \fr{1}{2m_{n+1}} \ga^0 \times {\rm Odd} (H_{n+1})
\quad .
\label{sj}
\eeq
Note that
\beq
H_{n+1} = \left\{ \prod_{k=0}^{n} \exp[\ad(iS_k)] \right\} H_0
\quad ,
\label{hj2}
\eeq
where the product represents map composition.

In the next section,
we find that each $H_{n+1}$ can be written in the form
\beq
H_{n+1}= m_{n+1} (\ga^0+\cp_{n+1}+\co_{n+1}+\ce_{n+1})
\quad ,
\label{hj}
\eeq
where the decomposition has the following useful properties:
(i) $\cp_{n+1}$ and $\co_{n+1}$ are odd;
(ii) $\ce_{n+1}$ is even;
(iii) $\co_{n+1}$ and $\ce_{n+1}$ are first order in parameters
for Lorentz violation;
and (iv) $\cp_{n+1}^2$ is proportional to the identity matrix,
with proportionality coefficient $t_{n+1}^2$
determined by $t_n^2$.
The existence of a decomposition of the form \rf{hj}
for arbitrary $n$,
as well as the case \rf{h0} above,
is a key feature making it feasible to calculate
the quantum particle hamiltonian.

\section{Calculation of the FW Sequence}
\label{calculation}

To calculate the FW sequence defined in the previous section,
the explicit form is needed of the operator
$\exp[\ad(iS_n)]$ connecting $H_n$ to $H_{n+1}$
according to Eq.\ \rf{hjplus1}.
Although $\ad(iS_n)H_n$ can be obtained directly 
using the properties of the Dirac matrices, 
calculation of $\exp[\ad(iS_j)]H_n$
is more challenging because
it is defined by an infinite series.
To address this issue,
we adopt the following approach: 
regard $\ad(iS_n)$ as a linear map
on a suitable vector space $\cv_n$ containing 
both $H_n$ and $H_{n+1}$,
and find a matrix expression of this map 
that can be exponentiated.

The first step in implementing this approach
is to define $\cv_n$ for each $n$.
It is convenient to introduce $\cv_n$ as the span
of a set of basis vectors $\cb_n$,
defined in terms of the operators
$\ga^0$, $\cp_n$, $\co_n$, $\ce_n$ determining $H_n$
together with the particular combinations of these four operators
that determine $\ad(iS_n)H_n$ and thus also $H_{n+1}$.
For each $n$, we define the ordered set 
\beq
\cb_n:= \left\{ \ga^0, \cp_n, \co_n, \cp_n\{\cp_n,\co_n\},
\ga^0[\cp_n,\ce_n],
   \ce_n, \ga^0\{\cp_n,\co_n\}, \cp_n[\cp_n,\ce_n] \right\}
\quad .
\label{B}
\eeq
The eight-dimensional vector space $\cv_n$ is formally defined 
as the real span of this set, 
so the elements of $\cb_n$ by definition form 
a (linearly independent) basis.
One advantage of this vector space is its 
relatively small dimensionality,
which makes it susceptible to practical calculation.
We can thus specify a vector $V \in \cv_n$ by eight
components $V_1,\ldots,V_8$:
\bea
V&:=& V_1 \ga^0 + V_2 \cp_n + V_3 \co_n + V_4 \cp_n \{\cp_n,\co_n\} 
  + V_5 \ga^0 [\cp_n,\ce_n] + V_6 \ce_n 
  + V_7 \ga^0 \{\cp_n,\co_n\}
  + V_8 \cp_n[\cp_n,\ce_n] \nonumber\\ 
 &\leftrightarrow& (V_1, \ldots ,V_8)   
\quad .
\label{G}
\eea
For example,
$H_n \leftrightarrow m_n(1,1,1,0,0,1,0,0)$.

The reader is warned to avoid confusing
the properties of the elements \rf{B}
as a basis for the vector space $\cv_n$ 
with their possible relationships when viewed 
as operators on the Hilbert space of wave functions.
For example,
the calculations below hold even if certain basis elements 
vanish as operators.
Note also that for different $n$
the corresponding vector spaces $\cv_n$ differ
\it a priori. \rm
However,
since both $H_n \in \cv_n$ and $H_{n+1} \in \cv_n$,
the vector space $\cv_n$ is invariant under
the action of $\exp[\ad(iS_n)]$,
which means $\cv_n \supseteq \cv_{n+1}$ for all $n$.

With the above notation,
we can present the results of a direct calculation
of $\ad(iS_n)V$ for $V \in \cv_n$ performed
using the properties of the Dirac matrices:
\beq
\ad(iS_n)V
\leftrightarrow (t_n^2 V_2,\ -V_1,\ -V_1,\ -V_7,\ 
     \half V_6+t_n^2 V_8,\
     0,\ \half V_2+\half V_3+t_n^2 V_4,\ -V_5 )
\quad .
\label{adiSG}
\eeq
In this expression,
$t_n^2$ is determined iteratively from $t_{n-1}^2$
through the relation
\beq
t_{n+1}^2= \left( \fr{ \cos{t_n}-\frac{\sin{t_n}}{t_n} }{ \cos{t_n}+t_n
\sin{t_n} } \right)^2 t_n^2
\quad .
\label{tk2}
\eeq
Here and in what follows,
we define functions of $t_n$ 
through their power-series expressions.
All relevant functions of $t_n$ implicitly involve 
only powers of $t_n^2$
(and hence powers of $t_0^2={\abs{\vec{p}}^2}/{m^2}$),
so it suffices to define $t_n^2$.
Note that $t_{n+1} \sim t_n^3$ to leading order in $t_n$,
so $t_n \sim t_0^{(3^n)}$.
This means that $t_n$ rapidly approaches zero
if $t_0 \ll 1$,
which ultimately is the reason for the rapid
convergence of our FW sequence.

With respect to the basis $\cb_n$,
the matrix map of $\ad(iS_n)$ can be extracted from
Eq.\ \rf{adiSG} and is given by
\beq
\ad(iS_n) \leftrightarrow \left(
 \begin{array}{cccccccc}
   0 & t_n^2 & 0 & 0 & 0 & 0 & 0 & 0  \\
   -1 & 0 & 0 & 0 & 0 & 0 & 0 & 0  \\
   -1 & 0 & 0 & 0 & 0 & 0 & 0 & 0  \\
   0 & 0 & 0 & 0 & 0 & 0 & -1 & 0  \\
   0 & 0 & 0 & 0 & 0 & \half & 0 & t_n^2  \\
   0 & 0 & 0 & 0 & 0 & 0 & 0 & 0  \\
   0 & \half & \half & t_n^2 & 0 & 0 & 0 & 0  \\
   0 & 0 & 0 & 0 & -1 & 0 & 0 & 0  \\
 \end{array}   \right)
\quad .
\label{arradiS}
\eeq
The exponential of this matrix can be found in closed form,
but its detailed expression is unimportant.
It can be used to calculate $\exp[\ad(iS_n)]H_n$,
which allows us to express $H_{n+1}$ in terms of $H_n$
according to Eq.\ \rf{hj} with
\bea
m_{n+1}&=& \left(\cos{t_n}+t_n \sin{t_n}\right) m_n 
\quad , \quad
m_{n+1}\cp_{n+1}= \left(\cos{t_n}-\fr{\sin{t_n}}{t_n}\right) m_n
   \cp_n 
\quad , 
\nonumber \\
m_{n+1}\co_{n+1}&=& \left(\cos{t_n}-\fr{\sin{t_n}}{t_n}\right) 
    m_n \co_n 
+ \fr{1}{2t_n^2}\left(\fr{\sin{t_n}}{t_n}-t_n\sin{t_n}-\cos{t_n}\right)
    m_n \cp_n \{\cp_n,\co_n\} 
+ \fr{\sin{t_n}}{2t_n} m_n \ga^0 [\cp_n,\ce_n] 
\quad , 
\nonumber \\
m_{n+1}\ce_{n+1}&=& m_n\ce_n 
 + \left(\half \cos{t_n}\right) m_n \ga^0 \{\cp_n,\co_n\} 
 + \left(\fr{\cos{t_n}-1}{2t_n^2}\right) m_n \cp_n[\cp_n,\ce_n]
\quad .
\label{mpoe2}
\eea

A measure of the convergence of the FW sequence
can be introduced using $t_n\sim t_0^{(3^n)}$.
In terms of a suitable matrix norm,
$\norm{{\rm Odd}(H_{n+1})}
\sim t_n^2\norm{{\rm Odd}(H_{n})} + t_n \norm{\ce_0}
\sim t_0^{2(3^n)}\norm{{\rm Odd} (H_{n})} + t_0^{(3^n)} \norm{\ce_0}$.
Thus, 
as $n$ grows
$\norm{{\rm Odd}(H_{n})}$ rapidly approaches zero 
as $({\abs{\vec{p}}}/{m})^{3^n}$.
Even a relatively small value of $n$ can therefore produce
a good approximation to the quantum particle hamiltonian.

\section{Nonrelativistic Quantum Hamiltonian}
\label{nrham}

The quantum particle hamiltonian $h_{\rm rel}$
and its nonrelativistic quantum limit $h$ are generated in
the limit of the FW sequence studied in the last section.
Next,
we demonstrate how to obtain these  
using simple matrix multiplication,
and we explicitly present $h_{\rm rel}$ and 
$h$ to order $t_0^3$. 

The calculation at the $k$th-iteration level 
in the FW sequence requires obtaining the composite map
$\prod_{n=0}^{k} \exp[\ad(iS_n)]$.
For each $n$ in the FW sequence,
the matrix $\ad(iS_n)$ and the action of $\exp[\ad(iS_n)]$
are given with respect to the basis $\cb_n$.
Since in general the vector space $\cv_n$ varies with $n$,
immediate calculation of $\prod_{n=0}^{k} \exp[\ad(iS_n)]$
by matrix multiplication is inappropriate.
Instead,
we first obtain the components of each matrix
with respect to the special basis $\cb_0$.
Ordinary matrix multiplication can then be used
to derive $\prod_{n=0}^{k} \exp[\ad(iS_n)]$.

The matrix for each map
$\ea{n}$ 
can be expressed in terms of $t_n$.
Explicitly, 
the nonzero entries for $\ea{n}$ 
with respect to the basis $\cb_0$ are:
\beq
\ea{n} \leftrightarrow
\left(
\begin{array}{cccccccc}
c_n+t_n^2 s_n & 0 & 0 & 0 & 0 & 0 & 0 & 0
\\
0 & c_n-s_n& 0 & 0 & 0 & 0 & 0 & 0
\\
0 & 0 & c_n-s_n & 0 & 0 & 0 & 0 & 0
\\
0 & 0 & (s_n-t_n^2s_n-c_n)/2t_0^2 & -t_n^2s_n & 0 & 0
& -t_ns_n/t_0 & 0
\\
0 & 0 & 0 & 0 & -t_n^2s_n & t_ns_n/2t_0 & 0 & t_0t_ns_n
\\
0 & 0 & 0 & 0 & 0 & 1 & 0 & 0
\\
0 & 0 & t_nc_n/2t_0 & t_0t_nc_n & 0 & 0 & c_n & 0
\\
0 & 0 & 0 & 0 & -t_nc_n/t_0 & (c_n-1)/2t_0^2 & 0 & c_n
\\
\end{array}
\right)
\quad ,
\label{ea_n2}
\eeq
where we have defined 
$c_n:=\cos{t_n} \approx 1-\frac{1}{2}t_n^2$
and $s_n:=\sin{t_n}/t_n \approx 1-\frac{1}{6}t_n^2$.

Since $t_n \rightarrow 0$ as $n \rightarrow \infty$,
it follows that $\ea{n}$ becomes a diagonal matrix
with entries $(1,0,0,0,0,1,1,1)$
in this limit.
The product $\prod_{n=0}^{k} \ea{n}$ 
therefore converges as $k \rightarrow \infty$,
so the limiting FW sequence giving the
quantum particle hamiltonian indeed exists. 
It can be shown that
\beq
\prod_{j=0}^{\infty}\ea{j}
\leftrightarrow \left(
\begin{array}{cccccccc}
  \ga &0&0&0&0&0&0&0 \\
  0&0&0&0&0&0&0&0 \\
  0&0&0&0&0&0&0&0 \\
  0&0&0&0&0&0&0&0 \\
  0&0&0&0&0&0&0&0 \\
  0&0&0&0&0&1&0&0 \\
  0&0& \fr{1}{2\ga} & [t_0^2] &0&0& [1-\half t_0^2] &0 \\
  0&0&0&0& [-1+\half t_0^2] & -\fr{1}{2\ga(\ga+1)} &0& [1-\half t_0^2] \\
\end{array}
\right)
\quad ,
\label{prodea}
\eeq
where $\ga:=\sqrt{1+t_0^2}$ is the usual relativistic gamma factor. 
Since the limiting FW hamiltonian is obtained 
by applying this matrix to 
$H_0 \leftrightarrow (1,1,1,0,0,1,0,0)$, 
the entries in brackets are irrelevant and
so we have evaluated them only to order $t_0^2$.
Thus,
we find the limiting FW hamiltonian to be
\beq
\tilde H
 = \ga m_0 \ga^0 + m_0 \ce_0
  + \fr{m_0}{2\ga} \ga^0 \{\cp_0,\co_0\}
  - \fr{m_0}{2\ga(\ga+1)} \cp_0[\cp_0,\ce_0]
\quad ,
\label{h3}
\eeq
an expression that is accurate to all orders in $t_0$.
Substitution from Eq.\ \rf{relham3}
yields the explicit form
\bea
\tilde H
&=& \ga m \ga^0
 + \Bigg\{ 
     a_0-me_0-m(c_{0j}+c_{j0})\fr{p^j}{m}
   \Bigg\}
+\Bigg\{
     -\fr{mc_{00}}{\ga}+(a_j-me_j)\fr{p^j}{\ga m}
     -m(c_{jk}-\et_{jk}c_{00})\fr{p^jp^k}{\ga m^2}
    \Bigg\} \ga^0
 \nonumber \\
&& +\Bigg\{
     -(md_{j0}+\half{\ve^{kl}}_{j}H_{kl})
     +\left[
       -\fr{b_0\et_{jk}}{\ga}
       +m{\ve^{lm}}_{j}(\half g_{lmk}-\et_{km}g_{l00})
      \right] \fr{p^k}{m}
 \nonumber \\
&& \qquad
    +\left[
       m(d_{0l}+d_{l0})
       -\fr{(\ga-1)m^2}{p^2}(md_{l0}+\half{\ve^{mn}}_{l}H_{mn})
     \right] \et_{jk}\fr{p^lp^k}{\ga m^2}
    +\left[
     \fr{(\ga-1)m^2}{2p^2}
     m{\ve^{nq}}_{l}g_{nqk}
    \right] \et_{jm}\fr{p^kp^lp^m}{\ga m^3}
   \Bigg\} \ga_5 \ga^j
 \nonumber \\
&& +\Bigg\{
     \left[-b_j-\half m{\ve^{kl}}_{j}g_{kl0}\right]\fr{1}{\ga}
     +\left[
       {\ve^{l}}_{kj}H_{0l}+m(d_{jk}-\et_{jk}d_{00})
      \right] \fr{p^k}{\ga m}
 \nonumber \\ 
&& \qquad 
    +\left[
     m{\ve^{m}}_{lj}(g_{m0k}+g_{mk0})
     +\fr{(\ga-1)m^2}{p^2}
      \et_{jl} (b_k+\half m{\ve^{mn}}_{k}g_{mn0})
    \right] \fr{p^kp^l}{\ga m^2} 
 \nonumber \\
&& \qquad 
    +\left[
      -\fr{(\ga-1)m^2}{p^2}
       m(d_{kl}-\et_{kl}d_{00})
    \right] \et_{jm}\fr{p^kp^lp^m}{\ga m^3}
  \Bigg\} \ga_5\ga^0\ga^j
\quad .
\label{FWham}
\eea   
This equation gives the FW form of the relativistic quantum hamiltonian
for a four-component fermion.

Certain limiting forms of the above expression 
are directly relevant to experiment. 
For applications involving relativistic two-component particles, 
such as the analysis of muon storage-ring experiments
\cite{bkl},
it suffices to retain only the upper left block $h_{\rm rel}$ 
of $\tilde H$:
\bea
h_{\rm rel}
&=& \ga m 
 + (a_0-\fr{mc_{00}}{\ga}-me_0)
     +\Big[
       a_j-\ga m(c_{0j}+c_{j0})-me_j       
     \Big] \fr{p^j}{\ga m}
     -m(c_{jk}-\et_{jk}c_{00})\fr{p^j p^k}{\ga m^2}
 \nonumber \\
&& +\Bigg\{
     \Big[
       -\fr{1}{\ga}b_j+md_{j0}+\half{\ve^{kl}}_{j}H_{kl}
       -\fr{1}{2\ga}m{\ve^{kl}}_{j}g_{kl0}
     \Big]
 \nonumber \\
&& \quad 
   +\Big[
    \et_{jk}b_{0}+m(d_{jk}-\et_{jk}d_{00})+{\ve^{l}}_{kj}H_{0l}
    -\ga m{\ve^{lm}}_{j}(\half g_{lmk}-\et_{km}g_{l00})
   \Big] \fr{p^k}{\ga m}
 \nonumber \\
&& \quad 
   +\Big[
     \fr{(\ga-1)m^2}{p^2}
     (b_k+md_{k0}+\half {\ve^{mn}}_{k}H_{mn}
       +\half m{\ve^{mn}}_{k}g_{mn0})
     \et_{jl}
 \nonumber \\
&& \qquad \qquad
    -m(d_{0k}+d_{k0})\et_{jl}
    +m{\ve^{m}}_{lj}(g_{m0k}+g_{mk0})
   \Big] \fr{p^k p^l}{\ga m^2}
 \nonumber \\
&& \quad 
   +\fr{(\ga-1)m^2}{p^2} 
   \Big[
    -m(d_{kl}-\et_{kl}d_{00}) 
    -\half m{\ve^{nq}}_{l}g_{nqk}
   \Big] \et_{jm} \fr{p^k p^l p^m}{\ga m^3}
  \Bigg\} \si^j
\quad .
\label{FW_ul}
\eea   
This is the quantum particle hamiltonian
associated with the original Lorentz-violating theory.

For many low-energy applications, 
including analyses of high-precision atomic experiments
\cite{bkr,gg,hd,rm,cptcc}, 
only nonrelativistic and subleading relativistic terms 
in the quantum particle hamiltonian are needed.
To third order in $|\vec p|/m$,
the nonrelativistic quantum hamiltonian $h$ for the two-component fermion is
\bea
h 
&=& m + \frac{p^2}{2m}
 \nonumber \\
&& +\left( a_0 -m c_{00} -m e_0 \right) 
+\left( -b_j + m d_{j0} - \half m \ve_{jkl}g_{kl0} 
      + \half \ve_{jkl}H_{kl} \right) \si^j  
+\left[-a_j+ m(c_{0j}+c_{j0}) +m e_j \right] \fr{p_j}{m} 
\nonumber \\ 
&& +\left[ b_0 \de_{jk} - m(d_{kj} +d_{00}\de_{jk}) 
       - m \ve_{klm}(\half g_{mlj}+g_{m00}\de_{jl}) -\ve_{jkl} H_{l0} 
       \right] \fr{p_j}{m} \si^k
+\left[ m(-c_{jk}-\frac{1}{2}c_{00}\de_{jk}) \right] 
     \fr{p_j p_k}{m^2} 
  \nonumber \\
&& +\left\{
       \left[ m(d_{0j}+d_{j0})
       -\half\left( b_j+md_{j0}+\half m\ve_{jmn}g_{mn0}
       +\half\ve_{jmn}H_{mn} \right)
       \right] \de_{kl} 
       \right .
      \nonumber \\
    && 
    \qquad \qquad \qquad \qquad
       \left .
       +\half\left(b_l+\half m\ve_{lmn}g_{mn0}\right)\de_{jk}
       -m \ve_{jlm} (g_{m0k}+g_{mk0}) 
       \right\} \fr{p_j p_k}{m^2}  \si^l 
  \nonumber \\
&& +\half\left( a_j\de_{kl}-me_j\de_{kl}\right)
       \fr{p_jp_kp_l}{m^3}
  \nonumber \\
&& +\half\left[
   \left( -b_0\de_{jm}+md_{mj}+\ve_{jmn}H_{n0}
    \right) \de_{kl}
      +\left( -md_{jk}-\half m\ve_{knp}g_{npj}
   \right) \de_{lm}
   \right]
   \fr{p_jp_kp_l\si^m}{m^3} 
   \quad .
\label{nrham3}
\eea 
Note that the form of Eq.\ \rf{h3} includes
all even elements of the basis set $\cb_0$.
This means that all possible combinations of the 
parameters for Lorentz violation
are already contained in Eq.\ \rf{nrham3}.
Higher-order corrections to the nonrelativistic hamiltonian
involve only products of these combinations with
powers of $|\vec p|^2/m^2$.
One interesting implication of this result
is that nonrelativistic experiments with single free fermions
(or fermions in weak external fields)
can at most be sensitive to the particular linear combinations
of parameters for Lorentz violation appearing in Eq.\ \rf{nrham3}.
Disentangling individual parameters
requires a different class of experiment.

As a final remark,
note that our methods can also be used to obtain
the nonrelativistic quantum hamiltonian $\overline{h}$ for the antifermion.
The result for $\overline{h}$ 
can be expressed in the same form as Eq.\ \rf{nrham3},
with the substitutions 
\a$\to \ol{a}_\mu = -$\a, 
\b$\to \ol{b}_\mu = +$\b, 
\c$\to \ol{c}_{\mu\nu} = +$\c, 
\d$\to \ol{d}_{\mu\nu} = -$\d, 
\e$\to \ol{e}_{\mu} = -$\e, 
\f$\to \ol{f}_\mu = -$\f, 
\g$\to \ol{g}_{\la\mu\nu} = +$\g, 
\H$\to \ol{H}_{\mu\nu} = -$\H.
This result is useful for experiments 
testing Lorentz symmetry with antimatter.

\section{Acknowledgments}

We thank R.\ Bluhm for discussions.
This work was supported in part
by the United States Department of Energy
under grant number DE-FG02-91ER40661.

\end{document}